\documentclass[12pt]{article}
\usepackage{graphicx,psfrag,epsf}
\usepackage{enumerate}
\usepackage{natbib}
\usepackage{bm}
\usepackage{amssymb,amsmath}

\newcommand{\blind}{0}

\newtheorem{Remark}{Remark}

\newtheorem{proposition}{Proposition}

\usepackage{color}

\definecolor{eGreen}{rgb}{.057, .549,.065}

\usepackage{soul} 

\begin{document}

\def\spacingset#1{\renewcommand{\baselinestretch}%
{#1}\small\normalsize} \spacingset{1}


\if0\blind
{
  \title{\bf A single-index  model with a surface-link 
   for optimizing individualized dose rules}
  \author{Hyung Park\thanks{
    This work was supported by National Institute of Health (NIH) grant 5 R01 MH099003.}, \hspace{.2cm} Eva Petkova, \hspace{.2cm} Thaddeus Tarpey  \hspace{.2cm}\\
    Division of Biostatistics, Department of Population Health, New York University\\
    and \\
    R. Todd Ogden \\
    Department of Biostatistics, Columbia University}
  \maketitle
} \fi

\if1\blind
{
  \bigskip
  \bigskip
  \bigskip
  \begin{center}
    {\LARGE\bf Title}
\end{center}
  \medskip
} \fi

\bigskip
\begin{abstract}
This paper focuses on the problem of modeling interaction effects between covariates and a continuous treatment variable on an outcome, using a single-index regression. The primary motivation is to estimate an optimal individualized dose rule 
in an observational study.  
To model possibly nonlinear interaction effects between  patients' covariates and a continuous treatment variable,  
we employ a two-dimensional penalized spline regression 
on an index-treatment domain, 
where the index is defined as a linear projection of the covariates. 
 The method is illustrated using two applications as well as simulation experiments. 
A unique contribution of this work is in the parsimonious (single-index) parametrization  specifically defined for the interaction effect term.

\end{abstract}

\noindent%

\spacingset{1.45}
\section{Introduction} \label{sec:intro}

In precision medicine, a primary goal is to characterize individuals' heterogeneity in treatment responses 
so that individual-specific treatment decisions can be made \citep{Murphy, Robins}. 
Most work on developing methods for individualized treatment decisions has focused on a finite 
number of treatment options. The focus of this paper is to develop individualized treatment decision methodology in the realm of a continuous treatment. Specifically, we consider a semiparametric regression approach for developing optimal individualized dosing rules based on baseline patient characteristics. 
Often in clinical practice, the maximum dose that a patient can tolerate is the most effective one, 
however, there are situations where this is not the case. 
In the example section, we present a study of warfarin (an anticoagulant), 
where too high doses lead to severe bleeding and thus the highest dose is not the the optimal dose. 
In finding the optimal dose, there is an essential non-monotone and nonlinear relationship that needs to be accounted for.  
A similar case is with insulin for controlling blood glucose levels. 

To establish notation, 
let  $X = ( X_1,\ldots, X_p)^\top \in \mathcal{X}$  be the set of baseline covariates, 
 $Y \in \mathbb{R}$ be the  outcome variable, and  $A \in \mathcal{A}$ denote the dose. 
Let $Y^\ast(a)$ be the potential outcome when a dose level $a \in \mathcal{A}$ is given. 
 Throughout, we assume:  
  1) the consistency, i.e., $Y = \int_{\mathcal{A}}\delta(A=a) Y^\ast(a) da$, where 
 $\delta(\cdot)$ is the Direc delta function; 2) the no unmeasured confoundedness, i.e., 
 $\{ Y^\ast(a), a \in \mathcal{A} \}$ are conditionally independent of $A$ given $X$; 
 3) the positivity, i.e., 
 $p(A=a | X=x) \ge c$, for all $a \in \mathcal{A}, x \in \mathcal{X}$, 
 for some $c >0$ (where $p(a|x)$ is the conditional density of $A=a$ given $X=x$), 
as in those adopted in the causal inference literature. 
Without loss of generality, we assume that a larger value of the outcome $Y$ is better. 
The goal is then to find an optimal individualized dose rule $f : \mathcal{X} \mapsto \mathcal{A}$ such that for a patient with covariate $X$, 
the dose assignment $A = f (X)$ maximizes the expected response, the so-called value function,  
$\mathcal{V}(f)  = \mathbb{E}[Y^\ast(f(X))]$, that is, 
\begin{equation} \label{value.function}
\mathcal{V}(f) = \mathbb{E}[ \mathbb{E}[Y | A=f(X), X ] ], 
\end{equation}
which holds and can be  empirically approximated under the above three assumptions.  
In settings in which the treatment can be administered at continuous doses (i.e., when $\mathcal{A}$ is an interval), \cite{Chen.IDR} proposed to 
optimize the individualized dosing rule $f$ 
by maximizing a local approximation of the value function (\ref{value.function}), optimized under the framework of outcome weighted learning \citep{Zhao.2012}, and 
\cite{Laber.Zhao.2015} proposed  
a tree-based decision rule for treatment assignment with minimal impurity dividing  patients into subgroups with different discretized doses. 
\cite{Kallus2018} 
developed an inverse propensity weighted estimator of (\ref{value.function}) for continuous treatments with the doubly robust property \citep{Dudik2015}, and recently, 
kernel assisted learning 
with linear dimension reduction 
 \citep{Zhou2020, Zhu2020} for direct optimization of (\ref{value.function}) 
  have been developed. However, 
  implementation of these approaches for 
  general  exponential family distributions is not straightforward and is not discussed. 
In this paper, we consider a regression-based approach to optimizing $f$ that uses a semiparametric regression model for $\mathbb{E}[Y | A,X]$. 
There is also extensive literature on multi-armed bandit \citep[e.g.,][]{Lattimore2019} problems  in the context of reinforcement learning \citep[e.g.,][]{Kaelbling1996},   
 incorporating  context (i.e.,   feature   $X \in \mathcal{X}$) (see, e.g., \cite{Lu2010, Perchet2013, Slivkins2014, Jun2017, Li2017, Kveton2020, Chen2020}) for making a sequential decision that minimizes the notion of cumulative regret, with relatively fewer works on contextual bandits with continuous actions \citep[see, e.g.,][]{Krishnamurthy2020, Kleinberg2019, Majzoubi2020}. 
 However, these works are focused on optimizing online performance (addressing the exploration issue) and considerably different from personalized dose-finding focused on a single stage with feature $X$.  \cite{Kennedy2017} considered a method  for estimating the average dose effect allowing for flexible doubly robust covariate adjustment, but the method is not intended for personalized dose-finding. For multi-stage personalized dose-finding, \cite{Rich2014} proposed adaptive strategies,  
and more recently \cite{Schulz2020} developed a doubly robust estimation approach based on a linear model. However, these approaches are limited by the stringent linear model assumptions for the heterogeneous dose effects. 
While the methods of directly optimizing the value function (\ref{value.function}), including the outcome weighted learning of \cite{Chen.IDR} and the tree-based method of \cite{Laber.Zhao.2015},  are highly appealing, the proposed semi-parametric 
regression modeling has the advantage of being easy to implement and readily generalizable to exponential family response.

It is straightforward to see that, given $X$, the optimal dose  $f_{\mathrm{opt}}(X)$ (i.e., that which maximizes the value function  (\ref{value.function})) is  
\begin{equation} \label{f.opt}
f_{\mathrm{opt}}(X)  \ = \   \underset{a \in \cal{A}}{\text{argmax}} \  m(a, X), 
\end{equation}
where $m(a, X) = \mathbb{E}[Y | A = a, X].$ 
If we estimate 
$m(a, X)$ with $\hat{m}(a, X)$, 
then the optimal rule $f_{\mathrm{opt}}$ in (\ref{f.opt}) can be approximated as
\begin{equation} \label{f.opt2}
\hat{f}(X)  \ = \   \underset{a \in \mathcal{A}}{\text{argmax}} \  \hat{m}(a, X).  
\end{equation}

Methodologies for optimizing individualized treatment rules $f$ in the precision medicine literature are  
mostly 
developed for the cases in which the treatment variable $A$  is binary or  discrete-valued.   
Regression-based methodologies  
first estimate the treatment $a$-specific mean response functions $m(a, X)$ and then obtain a treatment decision rule, i.e., 
the left-hand side of (\ref{f.opt2})  
\citep[e.g., see][]{QianAndMurphy, Zhang.classification, GUNTER.2011, LU.2011} given $X$.
In particular, 
 \cite{QianAndMurphy} show that the optimal individualized treatment rule (\ref{f.opt})   
depends only on the interaction  between treatment $A$ and covariates $X$, 
and not on the main effects for $X$ in the mean  models  $m(a, X)$.  
For regression-based methodologies, a successful estimation of the function $f_{\mathrm{opt}}$ in (\ref{f.opt}) boils down to efficiently estimating the $A$-by-$X$ interaction effects on the treatment response. In this paper, we consider a semi-parametric regression model that is
 useful for estimating such interactions in the case where $A$ is a continuous dose variable.

\section{Models}  \label{sec.models}

Our goal is to provide an interpretable and flexible approach to modeling the $A$-by-$X$ interaction effects on $Y$. 
To achieve this goal, we consider the following additive single-index model: 
\begin{equation}\label{the.truth} 
\mathbb{E}[Y | X, A]  \quad = \quad \mu(X)  \ + \   g(\beta^\top X, A)   
\end{equation}
where $\mu(X)$ represents an unspecified main effect of $X$, and $g(\beta^\top X, A)$ 
models 
 the $A$-by-$X$ interaction effects. Here, $g(\cdot, \cdot)$ is an unspecified smooth two-dimensional {\em surface} link function  
 of the variable 
$A$ and a single index $\beta^\top X$. 
We shall call model (\ref{the.truth}) a {\em single-index model with a surface-link} (SIMSL). 
We restrict $\beta \in \Theta := \{ \beta = (\beta_1, \ldots, \beta_p)^\top \in \mathbb{R}^p  :  \lVert \beta \rVert_2 = 1, \beta_1 >0 \}$,  
as $\beta$ in (\ref{the.truth}) is only identifiable up to a scale constant without further constraint, due to the unspecified nature of $g$. 

Without loss of generality,  we assume $\mathbb{E}[\mu(X) ] = 0$  and 
$\mathbb{E}[g(\beta^\top X, A)] = 0$ (where the expectation is with respect to $X$ and $A$), i.e., each of the additive components in model (\ref{the.truth}) has mean $0$, 
and that  
$\mu(X)$ and $g(\beta^\top X, A)$ have finite variances, as is typical assumed in 
  generalized additive models \citep[GAM;][]{GAM}.   
That is, let $\mathcal{H}_{1}$ and $\mathcal{H}_{2}^{(\beta)}$ (for a fixed $\beta \in \Theta$) 
denote the square-integrable spaces of measurable functions $\mu(X)$ on $X$ 
and measurable functions  $g(\beta^\top X, A)$ on $(\beta^\top X, A)$ (which depend on $\beta$), respectively, 
and we assume $\mu \in \mathcal{H}_{1}$ and  $g  \in \mathcal{H}_{2}^{(\beta)}$.

  To separate  the $A$-by-$X$ interaction effect of interest from the $X$ main effect in (\ref{the.truth}) and avoid confounding,  we 
will constrain the smooth function $g \in \mathcal{H}_{2}^{(\beta)}$ to satisfy: 
\begin{equation} \label{ortho.constr}
\mathbb{E}[g(\beta^\top X, A) | X]   
= 0, \quad   a.s.\ (X)  \quad X \in  \mathcal{X}, \   \beta \in \mathbb{R}^p, 
\end{equation} 
which acts as an identifiability condition of model (\ref{the.truth}).   
Applying the constraint (\ref{ortho.constr}) to the function $g$ in (\ref{the.truth}) essentially reparametrizes   
  the model (\ref{the.truth}), by replacing (centering) the component $g(\beta^\top X, A)$ 
with 
$g_0(\beta^\top X, A) = g(\beta^\top X, A) - \mathbb{E}[g(\beta^\top X, A) | X]$. 
The subtracted term $\mathbb{E}[g(\beta^\top X, A) | X]$ is added,  on balance,    to 
what was originally the $X$ ``main'' effect $\mu(X)$,  
replacing  $\mu(X)$ with  
$\mu_0(X) = \mu(X) + \mathbb{E}[g(\beta^\top X, A) | X]$.  
This yields an identifiable model of (\ref{the.truth}), $\mathbb{E}[Y | X, A] =  \mu_0(X)  +   g_0(\beta^\top X, A)$, 
 where the interaction function $g_0$ satisfies (\ref{ortho.constr}). Since any arbitrary $(\mu, g)$  in  (\ref{the.truth}) 
 can be rearranged to give such 
 reparametrized components 
 $(\mu_0, g_0)$,  
 we will   represent $(\mu_0, g_0)$ as  $(\mu, g)$ subject to (\ref{ortho.constr}).

Under the SIMSL (\ref{the.truth})  (subject to constraint (\ref{ortho.constr})),  
the optimal individualized dose rule, $f_{\mathrm{opt}}$, is specified as:  
$f_{\mathrm{opt}}(X) \ 
= \ \underset{a \in \mathcal{A}}{\text{argmax}} \  g(\beta^\top X, a)$, 
which does not involve the component $\mu$. 
Therefore,  in terms of estimating  $f_{\mathrm{opt}}$ in (\ref{f.opt}), our  modeling 
focus is on estimating $g$ and $\beta$ in (\ref{the.truth}).

Using the constrained least squares framework, 
the right-hand side of (\ref{the.truth}), subject to constraint (\ref{ortho.constr}), can be optimized by solving: 
 \begin{equation} \label{dddd}
\begin{aligned}
(\mu^\ast, g^\ast, \beta^\ast) \quad = \quad & \underset{\mu \in \mathcal{H}_{1}, g \in \mathcal{H}_{2}^{(\beta)}, \beta \in \Theta }{\text{argmin}}
& &\mathbb{E} \big[ (Y -  \mu(X) - g\big(\beta^{\top}X, A ))^2 \big]  \\
& \text{subject to} & &\mathbb{E}\left[g(\beta^\top X, A) | X \right] = 0.
\end{aligned}
\end{equation}

Constraint (\ref{ortho.constr}) ensures that 
 $\mathbb{E} \big[  \mu(X) g(\beta^\top X, A) \big]  =  \mathbb{E} \big[ \mu(X) \mathbb{E}[ g(\beta^\top X, A)|X] \big]= 0$ 
(in which we apply the iterated expectation rule to condition on $X$),  
 which implies the orthogonality, 
 \begin{equation} \label{ortho}
 \mu(X) \quad \perp \quad  g(\beta^\top X, A), 
  \end{equation} 
  in the $L^2$ space. 
The orthogonality (\ref{ortho}) implies that 
 the optimization for  
 $\mu^\ast$ and  that for $(g^\ast, \beta^\ast)$ on the left-hand side of   (\ref{dddd})  can be performed separately, without iterating between the two optimization procedures. 
Specifically,  we can solve for the $X$ main effect component: 
 \begin{equation}\label{sub.problem1}
\mu^\ast \quad = \quad \underset{\mu \in \mathcal{H}_{1}}{\text{argmin}} 
 \quad \mathbb{E} \big[ (Y -  \mu(X) )^2 \big], 
 \end{equation} 
and  separately  
 solve  for the $A$-by-$X$ interaction effect component:  
 \begin{equation} \label{sub.problem2}
\begin{aligned}
(g^\ast, \beta^\ast)  
 \quad = \quad  & \underset{g \in \mathcal{H}_{2}^{(\beta)}, \beta \in \Theta}{\text{argmin}}
& &\mathbb{E} \big[ 
(Y- g\big(\beta^{\top}X, A ))^2 \big]  \\
& \text{subject to} & &\mathbb{E}\left[g(\beta^\top X, A) | X \right] = 0. 
\end{aligned}
\end{equation}

The optimal individualized dose rule, $f_{\mathrm{opt}}$, is then  fitted as: 
$\hat{f}(X)   = \underset{a \in \mathcal{A}}{\text{argmax}} \quad  \hat{g}^\ast(\hat{\beta}^{^\ast \top} X, a)$, 
where  $(\hat{g}^\ast, \hat{\beta}^\ast)$ denotes 
an estimate of  
$(g^\ast, \beta^\ast)$  
on the left-hand side of 
(\ref{sub.problem2}).  
This optimization approach  (\ref{sub.problem2}) to estimating  $f_{\mathrm{opt}}$ in (\ref{f.opt}) is appealing, since, due to orthogonality (\ref{ortho}), 
  misspecification of the functional form for  $\mu$ in (\ref{dddd})  
(i.e., misspecification of $\mu^\ast$ in (\ref{sub.problem1})) 
   does not affect specification of $g^\ast$ and $\beta^\ast$ on the left-hand side of  (\ref{sub.problem2}). 
 If primary interest is in the $A$-by-$X$ interaction, 
as is generally 
the case when estimating $f_{\mathrm{opt}}$ in (\ref{f.opt}),  
we can thereby circumvent the need to estimate 
$\mu$ in (\ref{the.truth}),  
obviating the need to specify its form and thus avoiding the issue of model misspecification on the $X$ main effect.  
The equivalence between  $(g^\ast, \beta^\ast)$ on the left-hand side of (\ref{sub.problem2})  
and $(g, \beta)$ in (\ref{the.truth}) 
 is given in Proposition~\ref{proposition.1} of Section~\ref{sec.generalized}, 
in the context where $Y$ follows an exponential family response.

We focus on solving (\ref{sub.problem2}) as our primary focus is on  estimating  the $A$-by-$X$ interaction effect.  
For each fixed $\beta$, 
 the term $g(\beta^\top X, A)$ depends  the covariates $X \in \mathcal{X}$ only through the 1-dimensional projection $\beta^\top X$. 
  Therefore, for each fixed $\beta$,  
  the distribution of $g(\beta^\top X, A) | X $ 
is the same as that for  $g(\beta^\top X, A) | \beta^\top X$, 
indicating  
$\mathbb{E}\left[ g(\beta^\top X, A) |  X  \right]  =\mathbb{E}\left[ g(\beta^\top X, A) |  \beta^\top X  \right]$, 
for each fixed $\beta$. 
Then, 
for each fixed $\beta \in \Theta$, the following constraint on $g \in \mathcal{H}_2^{(\beta)}$,   
\begin{equation} \label{constraint2}
\mathbb{E}\left[ g(\beta^\top X, A) |  \beta^\top X  \right]   = 0,  \quad X \in \mathcal{X}, 
\end{equation}
is a  sufficient condition for the original ``orthogonality'' constraint (\ref{ortho.constr}). 
Thus, the original constraint (\ref{ortho.constr}) can be 
simplified to  (\ref{constraint2}), for each fixed $\beta$. 
The following iterative procedure will be used to solve (\ref{sub.problem2}):
\begin{enumerate}
\item
For fixed $\beta$, optimize the smooth $g(\cdot, \cdot)$ by solving: 
 \begin{equation}\label{sub.problem3}
\underset{g \in \mathcal{H}_{2}^{(\beta)}}{\text{argmin}} \ \mathbb{E} \big[ (Y - g\big(\beta^{\top}X, A ) )^2 \big], 
 \end{equation} 
subject to the constraint (\ref{constraint2}).  

\item
For fixed $g$, optimize the coefficient $\beta  \in \Theta$   by minimizing the  squared error criterion of (\ref{sub.problem3}). 
\item
Iterate steps (1) and (2) until convergence with $\beta \in \Theta$.
\end{enumerate}

The data version of optimizing $(g,\beta)$ can be derived as an empirical counterpart of the iterative procedure given above. Details on implementing this algorithm are given below.

\section{Estimation} \label{sec.estimation}

\subsection{Representation of link surface}  \label{sec.representation}
  Suppose we have observed data $(Y_i, A_i, X_i)$ $(i=1,\ldots,n)$.
For each candidate vector $\beta \in \Theta$, 
let 
$$
u_i = u_i^{(\beta)} = \beta^\top X_i \quad (i=1,\ldots,n), 
$$  
where (on the left-hand side),  for the notational simplicity,  we suppress   
the dependence of the linear predictor $u^{(\beta)} \in \mathbb{R}$ on the candidate vector $\beta$.

\cite{PSR.interaction} have used tensor products of $B$-splines \citep{deBoor}  to represent two-dimensional surfaces, which they termed  tensor product $P$-splines, with separate difference penalties applied to the coefficients of the $B$-splines along the covariate axes. 
    Although alternative nonparametric methods could also be used to estimate the 
 smooth function $g \in \mathcal{H}_2^{(\beta)}$ given each coefficient vector $\beta$ in model (\ref{the.truth}), 
    in this paper we focus on 
    one smoother, the tensor-product P-splines, for the ease of presentation.   

Specifically, for each $u = \beta^\top X$, 
to represent the $2$-dimensional function $g(u, A)$ 
 in (\ref{sub.problem3}),  
we consider the tensor product of the two sets of univariate cubic $B$-spline basis functions, 
say $B$ and $\check{B}$, 
with $N$ (and $\check{N}$) $B$-spline knots for the basis functions that are placed along the $u$ (and $A$) axis. 
The number of knots $N$ (and $\check{N}$) is chosen to be large, i.e., to allow the surface much flexibility. 
Associated with  the basis representation defined by the marginal basis function  
$B$ (resp., $\check{B}$)  
is an  $N \times N$ (resp., $\check{N}  \times \check{N}$) roughness penalty matrix,  
which we denote by $\mathbb{P}$ (and $\check{\mathbb{P}}$). 
The penalty matrix $\mathbb{P}$ (and $\check{\mathbb{P}}$) can be easily constructed, for example, based on a second-order difference matrix 
 (e.g, see \cite{PSR.interaction}). 
 
For each fixed $u_i = \beta^\top X_i$ $(i=1,\ldots,n)$,  
let us write the $n \times N$ (and $n \times \check{N}$) $B$-spline  evaluation matrix 
$\bm{B}$ (and $\check{\bm{B}}$), 
in which its $i$th row  is $\bm{B}_i = B(u_i)^\top$ (and $\check{\bm{B}}_i = \check{B}(A_i)^\top$). 
For a given knot grid, a flexible surface can be approximated \citep{VCSISR} at $n$ points $(u_i, A_i)$ $(i=1,\ldots, n)$: 
\begin{equation} \label{surface.splines}
g(u_i, A_i) 
=  \sum_{r=1}^{N} \sum_{s=1}^{\check{N}} B_{r}(u_i)  \check{B}_{s}(A_i) \gamma_{rs} 
 = (\bm{B}_i  \otimes  \check{\bm{B}}_i) \bm{\theta} 
\quad (i=1,\ldots,n), 
\end{equation}
where the vector 
$\bm{\theta} = \big( \gamma_{11}, \ldots, \gamma_{1\check{N}};  \ldots; \gamma_{N1}, 
 \ldots, \gamma_{N\check{N}}  \big)^\top  \in \mathbb{R}^{N \check{N}}$ corresponds to an unknown (vectorized) coefficient vector of the tensor product representation of $g$,  
and $\otimes$ represents the usual Kronecker product.  
Equation (\ref{surface.splines}) can be compactly written as:  
\begin{equation} \label{tensor.product}
\mbox{vec}\big\{  g(u_i, A_i)  \big\} = g( u_{n \times 1}, A_{n \times 1})  = \bm{D} \bm{\theta},
\end{equation}
where 
\begin{equation} \label{eq.D}
\bm{D} = \bm{B} \Box \check{\bm{B}} = \left( \bm{B} \otimes  \bm{1}_{\check{N}}^\top \right) \odot \left(  \bm{1}_{N}^\top  \otimes  \check{\bm{B}}  \right), 
\end{equation}
in which the symbol $\odot$ denotes element-wise multiplication of matrices.  
In \cite{Wood2017}, 
the symbol  $\Box$  in (\ref{eq.D}) is called 
the row-wise Kronecker product, 
which results in a $n \times N\check{N}$ tensor product design matrix $\bm{D}$ from the two marginal design matrices $\bm{B}$ and $\check{\bm{B}}$.

Similarly, 
the roughness penalty matrices associated with 
the tensor product representation (\ref{surface.splines})
can be constructed from 
the  roughness penalty matrices $\mathbb{P}$ and $\check{\mathbb{P}}$ 
associated with the univariate (marginal) basis matrices  $\bm{B}$ and $\check{\bm{B}}$, 
and are given by 
$\bm{P} = \mathbb{P} \otimes  \bm{I}_{\check{N}}$ and $\check{\bm{P}} = \bm{I}_{N} \otimes  \check{\mathbb{P}}$, 
for the axis directions $u$ and $A$, respectively. Here, 
$\bm{I}$ denotes the identity matrix, and both $\bm{P}$ and  $\check{\bm{P}}$
are square matrices with dimension $N \check{N}$.

We now need to impose the constraint (\ref{constraint2}) on  the $2$-dimensional smooth function 
$g$ under the tensor product representation  (\ref{tensor.product}). 
For each fixed $\beta$, 
the constraint (\ref{constraint2}) on $g$ amounts to 
excluding the main effect of $u = \beta^\top X$ from the function $g$. 
We deal with this by a reparametrization of the representation (\ref{tensor.product}) for $g$.

Consider the following sum-to-zero (over the $n$ observed values) constraint for the marginal function of $A$: 
\begin{equation}\label{lin.constr} 
\bm{1}^\top \check{\bm{B}} \check{\bm{\gamma}} = 0,
\end{equation} 
for any arbitrary $\check{\bm{\gamma}} \in \mathbb{R}^{\check{N}}$, 
where $\bm{1}$ is a length $n$ vector of 1's.  
With constraint (\ref{lin.constr}), the linear smoother associated with the basis matrix $\check{\bm{B}}$ cannot reproduce constant functions \citep{GAM}.  
That is, 
the linear constraint (\ref{lin.constr}) removes the span of constant functions  
 from the span of the marginal basis matrix  $\check{\bm{B}}$ associated with $A$. 
 Constraint (\ref{lin.constr}) results in a tensor product basis  matrix, 
 $\bm{D} = \bm{B} \Box \check{\bm{B}}$ in (\ref{tensor.product}),  
  that will not include the main effect of $u$  
   that results from the product of the marginal basis matrix $\bm{B}$ 
  with the constant  function in the span of  
  the other marginal basis matrix $\check{\bm{B}}$. Therefore,  
 the resultant fit, under representation (\ref{tensor.product}) (subject to (\ref{lin.constr})) of the smooth function $g$, excludes the main effect of $u$. 
  See Section 5.6 of \cite{Wood2017} for some more details.

  We impose the linear constraint (\ref{lin.constr}) on the matrix $\check{\bm{B}}$, 
  and consequently, the resulting  basis matrix $\bm{D}$ of representation of $g$ in (\ref{tensor.product}) 
  becomes independent of the basis associated with the main effect of $u$. 
Imposition of such a linear constraint (\ref{lin.constr}) on a basis matrix is routine. 
The key is to find an (orthogonal) 
   basis  for the null space of the constraint  (\ref{lin.constr}), 
  and then 
 absorb the constraint into the basis construction  (\ref{eq.D}). 
 To be specific, 
 we can create a $\check{N} \times (\check{N}-1)$ matrix, which we denote as $\bm{Z}$, such that, 
 given any arbitrary coefficient vector $\check{\bm{\gamma}}_0 \in \mathbb{R}^{\check{N}-1}$, 
  if we set $\check{\bm{\gamma}} = \bm{Z} \check{\bm{\gamma}}_0$, then we have 
$\bm{1}^\top \check{\bm{B}} \check{\bm{\gamma}} = 0$, and thus automatically satisfy the constraint (\ref{lin.constr}). 
Such a  matrix 
$\bm{Z}$ is constructed using a QR decomposition of $\check{\bm{B}}^\top \bm{1}$. 
Then we can reparametrize the marginal function of $A$ by setting its model matrix to $\check{\bm{B}} \leftarrow  \check{\bm{B}} \bm{Z}$ 
(and its penalty matrix to $\check{\mathbb{P}} \leftarrow \bm{Z}^\top \check{\mathbb{P}} \bm{Z}$). 
From this point forward, for notational simplicity, we redefine the matrix $\check{\bm{B}}$ (and $\check{\mathbb{P}}$) to be this reparameterized, constrained marginal basis matrix (and the reparameterized  constrained penalty matrix).

This sum-to-zero reparametrization of the marginal basis matrix $\check{\bm{B}}$ of $A$ to satisfy (\ref{lin.constr}) is simple and creates a 
term $\mbox{vec}\big\{  g(u_i, A_i)  \big\} \in \mathbb{R}^n$   in (\ref{tensor.product}) that specifies such a pure $A$-by-$X$ interaction (plus the $A$ main effect) component, that is also  
orthogonal to the $X$ main effect.  
  In \cite{Wood2006}, this reparameterization approach is used to create an analysis of variance (ANOVA) decomposition of a smooth function of several variables.  In this paper we use  this same reparameterization to orthogonalize the interaction effect component $g(\beta^\top X, A)$ from the main effect, and to allow an unspecified/misspecified  
  main effect for $X$ in the estimation of the SIMSL (\ref{the.truth}). 
Provided that the orthogonality constraint (i.e., (\ref{lin.constr})) issue is addressed, the interaction effect term $g(\beta^\top X, A)$ of model (\ref{the.truth}), for each fixed $\beta$, 
can be represented using penalized regression splines and estimated based on penalized least squares, which we describe next.

\subsection{Estimation algorithm} \label{sec.estimation.algo}

We define the criterion function for estimating $(g, \beta)$ 
 in the SIMSL (\ref{the.truth}): 
\begin{equation} \label{Q.criterion}
\begin{aligned}
Q(\bm{\theta}, \beta) \ 
&= \ \lVert Y_{n \times 1} - g( \bm{X} \beta, A_{n \times 1}) \rVert^2 + \lambda \lVert \bm{P}\bm{\theta} \rVert^2  + \check{\lambda}  \lVert \check{\bm{P}} \bm{\theta}  \rVert^2 
\\ 
&=
\ \lVert  Y_{n \times 1}  - \bm{D} \bm{\theta} \rVert^2 + \lambda \lVert \bm{P}\bm{\theta} \rVert^2  + \check{\lambda}  \lVert \check{\bm{P}} \bm{\theta}  \rVert^2 
\end{aligned}
\end{equation} 
subject to the constraint that the function $g(\cdot, \cdot)$ empirically satisfies  (\ref{ortho.constr}). 
In (\ref{Q.criterion}), $\bm{X}$ is a $n \times p$ matrix whose $i$th row is $X_i^\top$.  
Since both $\bm{\theta}$ and $\beta$  are unknown in (\ref{Q.criterion}), 
estimation of $\bm{\theta}$ and $\beta$ is conducted iteratively. 
We describe below the estimation procedure.

\begin{enumerate}
\item
For a fixed estimate of $\beta$ (that defines the linear predictor $u$), 
minimize the following criterion function over $\bm{\theta} \in \mathbb{R}^{N\check{N}}$, 
\begin{equation} \label{Q.criterion2}
 \lVert  Y_{n \times 1}  - \bm{D} \bm{\theta} \rVert^2 + \lambda \lVert \bm{P}\bm{\theta} \rVert^2  + \check{\lambda}  \lVert  \check{\bm{P}} \bm{\theta}  \rVert^2,  
\end{equation} 
where $\bm{D}$ is given by (\ref{eq.D}). 
Given tuning parameters $(\lambda, \check{\lambda})$, 
the minimizer $\hat{\bm{\theta}}$ of (\ref{Q.criterion2}) is: 
$$
\hat{\bm{\theta}} = \left( \bm{D}^\top \bm{D} + \lambda \bm{P}^\top \bm{P} + \check{\lambda} \check{\bm{P}}^\top \check{\bm{P}}  \right)^{-1} \bm{D}^\top Y_{n \times 1}. 
$$ 

\item
For a fixed estimate of the surface $g$ (i.e., given $\bm{\theta}$), 
perform a first-order Taylor approximation of 
$g( \bm{X} \beta, A_{n \times 1}) $
 in (\ref{Q.criterion})
 with respect to $\beta$, around the current estimate, denote as $\tilde{\beta} \in \Theta$, 
\begin{equation}\label{taylor.approx}
g( \bm{X} \beta, A_{n \times 1}) \  \approx \ g( \bm{X}\tilde{\beta}, A_{n \times 1}) + \mbox{diag}\big\{ \dot{g}_{\partial_1}(\bm{X} \tilde{\beta}, A_{n \times 1})  \big\} \bm{X}(\beta - \tilde{\beta}),
\end{equation}
where 
$\dot{g}_{\partial_1 }(u,a)$ denotes 
the partial first derivative of $g(u,a)$ with respect to the first variable $u$, i.e.,  
$\frac{\partial g(u,a)}{\partial u}$.  
Utilizing (\ref{taylor.approx}), 
the quadratic loss term in (\ref{Q.criterion}), as a function of $\beta$ given $\bm{\theta}$, can be approximated as:  
\begin{equation}\label{Q.alpha}
\begin{aligned}
&  \  \left\lVert Y_{n \times 1} -  g( \bm{X} \tilde{\beta}, A_{n \times 1}) - \mbox{diag}\big\{ \dot{g}_{\partial_1 }(\bm{X} \tilde{\beta}, A_{n \times 1})  \big\} \bm{X}(\beta - \tilde{\beta})  \right\rVert^2  \\  
=&  \ 
\left\lVert 
 Y_{n \times 1}-  g( \bm{X} \tilde{\beta}, A_{n \times 1}) + \mbox{diag}\big\{ \dot{g}_{\partial_1}(\bm{X}\tilde{\beta}, A_{n \times 1})  \big\} \bm{X} \tilde{\beta} 
  -  \mbox{diag}\big\{ \dot{g}_{\partial_1 }(\bm{X} \tilde{\beta}, A_{n \times 1})  \big\} \bm{X} \beta  \right\rVert^2  \\ 
=&  \ 
\left\lVert  Y_{n \times 1}^\ast -  \bm{X}^\ast \beta  \right\rVert^2,
\end{aligned}
\end{equation}
where
$Y_{n \times 1}^\ast =  Y_{n \times 1} -  g( \bm{X} \tilde{\beta}, A_{n \times 1}) + \mbox{diag}\big\{ \dot{g}_{\partial_1 }(\bm{X} \tilde{\beta}, A_{n \times 1})  \big\} \bm{X} \tilde{\beta}$,  
and 
$\bm{X}^\ast = \mbox{diag}\big\{ \dot{g}_{\partial_1 }(\bm{X}\tilde{\beta}, A_{n \times 1})  \big\} \bm{X}$. 
The minimizer $\hat{\beta}$ of  (\ref{Q.alpha}) is: 
\begin{equation} \label{LS.beta}
\hat{\beta}=  \left( \bm{X}^{\ast \top} \bm{X}^\ast  \right)^{-1} \bm{X}^{\ast \top} Y_{n \times 1}^\ast. 
\end{equation} 
Then we scale $\hat{\beta}$ to  unit $L^2$ norm, i.e., 
$\hat{\beta} / \lVert \hat{\beta} \rVert$,  and enforce a positive first element to restrict the estimate of $\beta$ to be in $\Theta$. 
\end{enumerate}

These two steps can be iterated until convergence to obtain an estimate of $(g^\ast, \beta^\ast)$ in (\ref{sub.problem2}), 
which we denote as $(\hat{g}^\ast, \hat{\beta}^\ast)$. 
For Step 1, the tuning parameters $(\lambda, \check{\lambda})$ 
can be automatically selected, for example, by the generalized cross-validation (GCV) or the restricted maximum likelihood (REML) methods. 
In this paper, we use REML 
for the simulation examples and the applications. 

Lastly, for model hierarchy, it is common practice to include all lower order effects of variables if there are higher-order interaction terms including that set of variables. 
Once convergence of the estimate $\hat{\beta}^\ast$ is reached in the above algorithm  and  the single-index $\beta^\top X$ in the term $g(\beta^\top X, A)$ of model (\ref{the.truth}) is estimated,  
we recommend fitting one final (unconstrained)  
 smooth function $g$ of $A$ and $\hat{\beta}^{\ast\top} X$, without enforcing the constraint (\ref{lin.constr}) on $g$. 
Given the final estimate of $\beta$, 
 the  unconstrained final surface-link $g(\cdot,\cdot)$ retains the main effect of $\beta^\top X$ and preserves model hierarchy.

\section{  
Generalized single-index models for optimizing dose rules}  
\label{sec.generalized}

The proposed approach to optimizing the heterogeneous dose effect (i.e., the $X$-by-$A$ interaction effect)  
term of model (\ref{the.truth}) can be extended to a more general setting in which the response $Y$ follows an exponential family distribution.  We again assume an additive single-index model (\ref{the.truth}) for the true mean response function given $X \in \mathcal{X}$  and $A \in \mathcal{A}$:  
\begin{equation}\label{the.truth2} 
m(X,A)  =  \mathbb{E}[Y | X, A]  \ = \ \mu_0(X)  \ +  \  g_0(\beta_0^{\top} X, A),  
\end{equation} 
and the variance of $Y$ given $(X,A)$ is determined based on the exponential family density of the form: 
\begin{equation}\label{the.density}
\exp\left\{ \big[ Y h(m(X,A))  -  b( h(m(X,A)) ) \big]  / a(\phi)  + c(Y, \phi ) \right\},
\end{equation} 
where $h$ is the \textit{canonical} link  function 
associated with the assumed distribution of $Y$,  
 and the functions $a$, $b$ and $c$ are  
 distribution-specific  known  
functions. 
 In (\ref{the.truth2}), we used subscript $(0)$ to indicate the ``true'' value.  For model identifiability, we assume 
$\beta_0 \in \Theta$, and  
$g_0 \in \mathcal{H}_{2}^{(\beta_0)}$ to satisfy 
 $\mathbb{E}[g_0(\beta_0^{\top} X, A) | X ] = 0$.  
 The dispersion parameter $\phi > 0$ 
in (\ref{the.density}) takes on a fixed, known value  in some families (e.g., $\phi = 1$, in Bernoulli and Poisson), while in other families, it is an unknown parameter (e.g., in Gaussian).

Our approach to estimating $g_0(\beta_0^{\top} X, A)$ in model (\ref{the.truth2})  is to  utilize the following (misspecified) working model: 
 \begin{equation}\label{the.working.model} 
m(X,A) 
\ =  \    h^{-1}(g( \beta^\top X, A)) \quad \quad (\beta \in \Theta; \ g \in \mathcal{H}_{2}^{(\beta)}),   
\end{equation}
subject to the constraint 
$\mathbb{E}[g(\beta^\top X, A) | X ] = 0$ on $g$, with the exponential family distribution chosen in (\ref{the.density}). 
We propose to optimize the population version of the log  of the likelihood (\ref{the.density})
over the unknowns  $(g, \beta)$  of (\ref{the.working.model}):    
 \begin{equation} \label{glm.simsl.criterion}
\begin{aligned}
(g^\ast, \beta^\ast)  \quad = \quad  & \underset{g \in \mathcal{H}_{2}^{(\beta)}, \beta \in \Theta}{\text{argmax}}
& &\mathbb{E} \big[  Y g( \beta^\top X, A)  -  b(g(\beta^\top X, A) ) \big]  / a(\phi)  \\
& \text{subject to} & &\mathbb{E}\big[g(\beta^\top X, A) | X \big] = 0 
\end{aligned}
\end{equation}
where the expectation on the first line  is with respect to the joint distribution of $(Y, A, X)$.  
Thus, 
the solution $(g^\ast, \beta^\ast)$ on the left-hand side of (\ref{glm.simsl.criterion}) is  defined as the minimizer of the Kullback-Leibler (KL) divergence between the working model (\ref{the.working.model})  and the true model (\ref{the.truth2}).  
(Note that the scaling factor, $1 / a(\phi)$, in (\ref{glm.simsl.criterion}) can be dropped  if our interest is only in fitting the working mean model (\ref{the.working.model})). 
In (\ref{glm.simsl.criterion}), $b(s) = s^2/2$ for a Gaussian $Y$ (for which the optimization  (\ref{sub.problem2}) is a special case of (\ref{glm.simsl.criterion})), 
$b(s) = \log\{1+\exp(s)\}$ for a Bernoulli $Y$, 
and $b(s) = \exp(s)$ for a Poisson $Y$.   

 The constraint $\mathbb{E}\big[g(\beta^\top X, A) | X \big] = 0$ in  optimization  (\ref{glm.simsl.criterion}) implies: 
\begin{equation} \label{eq.for.Q2}
\begin{aligned}
\mathbb{E} \left[Y g(\beta^\top X, A)   -  b(g(\beta^\top X, A) )    \right]  
&=
\mathbb{E} \left[ \big\{\mu_0(X) + g_0(\beta_0^{\top} X, A) \big\} g(\beta^\top X, A)   -  b(g(\beta^\top X, A) )   \right]  \\
&=
\mathbb{E} \big[ \mathbb{E} \big[\mu_0(X) g(\beta^\top X, A)  | X \big] \big] +\mathbb{E} \big[  g_0(\beta_0^{\top} X, A) (g(\beta^\top X, A)  -  b(g(\beta^\top X, A) )   \big] \\
&= 
\mathbb{E} \big[  g_0(\beta_0^{\top} X, A) (g(\beta^\top X, A)  -  b(g(\beta^\top X, A) )   \big], \\
\end{aligned}
\end{equation}
which is free of the term $\mu_0(X)$ given in model  (\ref{the.truth2}). 
Therefore, the left-hand side, ($g^\ast, \beta^\ast$),  
of (\ref{glm.simsl.criterion})  
does not depend on the unspecified $X$ ``main'' effect term  $\mu_0(X)$ in (\ref{the.truth2}).  

 \begin{proposition} \label{proposition.1}
The solution 
$(g^\ast, \beta^\ast)$  of the constrained optimization problem (\ref{glm.simsl.criterion}) satisfies: 
\begin{equation}\label{the.solution}
g_0 = h^{-1} \circ g^\ast \quad \mbox{and} \quad \beta_0=  \beta^\ast,
\end{equation} 
where $g_0 \in \mathcal{H}_{2}^{(\beta_0)}$ and $\beta_0 \in \Theta$ are given from the true mean model (\ref{the.truth2}), 
and the function
  $h^{-1}$ is the inverse of the canonical link function associated with the assumed exponential family distribution  
and the operator $\circ$ represents  the composition of two functions.
  \end{proposition} 
  The proof of Proposition~\ref{proposition.1} is in Supplemental Materials. 
In (\ref{the.solution}),  $h^{-1}(s) = s$  (the identity function) 
 for a Gaussian $Y$,  
$h^{-1}(s) = \exp(s)/\{1+\exp(s)\}$ 
for a Bernoulli $Y$,  
and  $h^{-1}(s) =  \exp(s)$ for a Poisson $Y$.  
 
  In practice, to solve  (\ref{glm.simsl.criterion})  based on observed data $(Y_i, A_i, X_i)$ $(i=1,\ldots,n)$,  
we replace  the squared error term in  
(\ref{Q.criterion}) by the negative of the log likelihood of the data.  
For a fixed $\beta \in \Theta$, 
given smoothing parameter
values  ($\lambda$ and $\check{\lambda}$), and 
upon the  penalized spline basis function expansion for $g$ (i.e., the representation with $\bm{D}\bm{\theta}$ in (\ref{tensor.product})), 
the basis coefficient $\bm{\theta}$  
is estimated by the inner  iteratively re-weighted least squares (IRLS); the smoothing parameters are estimated
by the outer optimization of, for example,  REML or GCV  (see, for example, \cite{Wood2017}),  
 as part of Step 1 (of Section~\ref{sec.estimation.algo}) of the model fitting.    
The only adjustment  to be made to the conventional GAM  
is to enforce the constraint $\mathbb{E}[g(\beta^\top X, A) | \beta^\top X ] = 0$ on the smooth $g$. 
As in Section~\ref{sec.representation}, this constraint can be absorbed into the tensor product basis representation (\ref{tensor.product}).  
For Step 2 (in Section~\ref{sec.estimation.algo}) of the estimation,  once we profile out $g$ (by Step 1), we replace 
the Gaussian \textit{residual} vector   in (\ref{Q.alpha}), i.e.,  
$Y_{n \times 1} -  g( \bm{X} \tilde{\beta}, A_{n \times 1})$, where  $\tilde{\beta}$  denotes the estimate of $\beta$ from the previous iteration,    
 by  the \textit{working residual}  from the final IRLS  fit of Step 1,  and perform a weighted least squares (instead of the least squares (\ref{LS.beta}))  for $\beta$,  where the weights  
are given  from  the final IRLS fit of Step 1. 
The estimation   alternates  
between the Steps 1  
and  2 until convergence of $\hat{\beta}$, as in Section~\ref{sec.estimation.algo}.  The resulting estimate for $(g^\ast, \beta^\ast)$   in (\ref{glm.simsl.criterion}) is then used to estimate  
$(g_0, \beta_0)$ in (\ref{the.truth2}), based on the relationship (\ref{the.solution}).

\begin{Remark}
The proposed approach (\ref{glm.simsl.criterion}) 
to optimizing $(g, \beta)$ of SIMSL 
can be generalized  
 to the context of  
a  proportional odds  single-index 
model. 
Suppose  we have  
   an ordinal categorical variable $Y$,  
where its value exists on an arbitrary scale (in $K$ categories, say), with only the relative ordering between different values being important.    
To deal with such a case, we introduce  a  length-$K$ response  vector 
  $\bm{Y} = (Y_{1}, Y_{2},\ldots, Y_{K})^\top$,   
where its component  $Y_{j}$ denotes the indicator for category $j$, 
and the associated vector of probabilities  $(p_1, p_2,\ldots, p_K)^\top$,   
in which $p_K = 1 - \sum_{j=1}^{K-1}p_j$ (and $p_j > 0$), 
together with   their cumulative probabilities: $P(Y \le j)= q_j 
 = \sum_{s=1}^{j} p_s$  $(j=1,2,\ldots,K-1)$.  
We can model these cumulative probabilities  
 $(q_1, q_2, \ldots, q_{K-1}, 1)^\top$ 
by cumulative logit SIMSL: 
\begin{equation} \label{prop.odds}
h(P(Y \le j | X,A))  = 
 h(q_j(X,A)) 
 \ = \ 
  \alpha_j +  \mu(X) + 
   g(\beta^\top X, A) \quad (j=1,2,\ldots,K-1)  
 \end{equation}
 where  $\alpha_j \in \mathbb{R}$  $(\alpha_1 < \alpha_2 < \ldots < \alpha_{K-1})$ are  unknown cut-point parameters 
 associated with the ordered response categories  $j =1,2,\ldots,K-1$, 
 and  $h(s) = \log(s/(1-s))$ is the logit link. 
 For the ($K$-category) multinomial response $\bm{Y}$, 
let us consider its canonical parameter  $\bm{\eta} := (\eta_1, \eta_2,\ldots, \eta_{K-1},0)^\top \in \mathbb{R}^K$ where its nonzero components are specified by: $\eta_ j =  \log\big(p_j/p_K\big)  = \log\big((q_{j} - q_{j-1})/(1-q_{K-1})\big)$  $(j=1,\ldots,K-1)$ (with $q_{0}  = 0$), in which 
 the cumulative probabilities 
 $(q_1, q_2, \ldots, q_{K-1}, 1)^\top$  
are  specified by model  (\ref{prop.odds}). This multinomial exponential family representation for the distribution of $Y$  
 allows us to use the optimization framework (\ref{glm.simsl.criterion}),  
with its criterion function  extended to incorporate the multivariate response:  
 $\mathbb{E} \big[  \bm{Y}^\top \bm{\eta} -  b(\bm{\eta}) \big]$, 
where 
  $b(\bm{\eta}) = \log\{1 + \sum_{j=1}^{K-1} \exp(\eta_j) \}$, 
for  optimization of the cumulative logit SIMSL  (\ref{prop.odds}).  
The   threshold-point parameters 
$\alpha_j \in \mathbb{R}$ $(\alpha_1 < \alpha_2 < \ldots < \alpha_{K-1})$ in model (\ref{prop.odds}) are estimated as part of Step 1 of model fitting  in Section~\ref{sec.estimation.algo} (alongside the model smoothing parameters), 
in which the model (\ref{prop.odds}), for each fixed $\beta$ which is estimated as part of Step 2, is optimized via the penalized IRLS with an empirical version of $-\mathbb{E} \big[  \bm{Y}^\top \bm{\eta} -  b(\bm{\eta}) \big]$; the Steps 1 and 2 are iterated until convergence. 
The heterogeneous treatment effect $g(\beta^\top X, A)$ in model (\ref{prop.odds}) does not depend on $j$ nor $\mu(X)$; 
this allows us to develop an individualized dose rule  independently of the arbitrary categorization of $Y$, and of the unspecified $X$ main effect $\mu(X)$ that does not influence the treatment effect. 
For a patient with covariates $X$, 
the cumulative logit SIMSL-based individualized dose rule is $f(X) =  \underset{a \in \cal{A}}{\text{argmax}} \ g(\beta^\top X, a)$. 
 \end{Remark}

   In Supplemental Materials Section B, 
we provide a real data example illustrating the  approach (\ref{glm.simsl.criterion}) to modeling interaction effects
  between  $X$ and $A$ on a count response variable 
 $Y$, and a simulation example 
illustrating 
the utility of 
the proportional odds model (with a surface-link) (\ref{prop.odds})  in  optimizing dose $A$ based on features $X$, 
 when the treatment response $Y$ is  ordinal and categorical, which is common in biomedical/epidemiological studies and social sciences.

\section{Simulation example} \label{sec.simulation}

In this section, 
we consider a set of simulation studies with data generated from the four scenarios described in \cite{Chen.IDR}.  
We generate $p$-dimensional covariates  $X= (X_1,\ldots, X_p)^\top$, 
where each entry is generated 
independently from 
 $\mbox{Uniform}[-1,1]$.
 In Scenarios 1 and 2, 
   the treatment $A$ is generated from  $\mbox{Uniform}[0,2]$ independently of $X$, mimicking a randomized trial. 
    In Scenarios 3 and 4, the distribution of $A$  (described below) depends on $X$, mimicking an observational study setting. 
In each scenario,  the outcome $Y$, given $X$ and $A$, is generated from 
the standard normal distribution,  with the following four different mean function scenarios:

\begin{enumerate}
\item \ul{Scenario 1:}
$
\mathbb{E}[Y | X,A] =  8 + 4 X_1 - 2 X_2 - 2X_3 - 25 ( f_{\mathrm{opt}}(X) - A )^2,  
$
where 
$f_{\mathrm{opt}}(X) = 1 + 0.5 X_1 + 0.5  X_2.$ Here, the optimal  individualized dose rule  is a linear function of $X$. 

\item  \ul{Scenario 2:}
$
\mathbb{E}[Y | X,A]  =  8 + 4 \cos(2\pi X_2)  - 2 X_4  -  8X_5^3  -  15  |  f_{\mathrm{opt}}(X) - A |,  
$
where 
$$f_{\mathrm{opt}}(X) = 0.6(-0.5 < X_1 <  0.5) + 1.2 (X_1 > 0.5) + 1.2 (X_1 < -0.5) + X_4^2 +  0.5 \log( |X_7 | + 1)  - 0.6.$$
 Here, the optimal individualized dose rule is a nonlinear function of $X$. 

\item 
\ul{Scenario 3 } is the same as in Scenario 2, except that the distribution of $A$ depends on $X$ as follows: 
$$
A \sim \left\{
\begin{array}{ll}
\text{TruncN} \left( -0.5+0.5 X_{1}+0.5 X_{2}, 0,2,0.5 \right),  & \text{ when } X_{3}<0 \\
 \text { TruncN} \left( \left| 0.5+1.5 X_{2} \right|, 0,2,1 \right),  & \text{ when } X_{3}>0 \end{array}\right.
 $$
where 
$\text{TruncN}\left( \mu, a, b, \sigma \right) $ 
denotes the truncated normal distribution with mean $\mu$, lower bound $a$ and upper bound $b$, and standard deviation $\sigma$. 

\item  \ul{Scenario 4} is the
same as in Scenario 2, except that the distribution of $A$ depends on $X$ as follows: 
$$
A \sim \operatorname{TruncN}\left(f_{\mathrm{opt}}(X), 0,2,0.5\right). 
$$
\end{enumerate}

\begin{Remark}
We briefly describe how the  above data generation scenarios are related to  SIMSL (\ref{the.truth}) (subject to (\ref{ortho.constr})).  
For Scenario 1, by introducing  $\beta := (0.5, 0.5, 0, 0, \ldots, 0)^\top/\sqrt{0.5} \in \Theta$, we can write 
$f_{\mathrm{opt}}(X)   = 1  + \sqrt{0.5}  \beta^\top X$. 
On the other hand, under SIMSL (\ref{the.truth}), 
the term $g$ 
  is $g( \beta^\top X, A) :=  - 25 ( f_{\mathrm{opt}}(X) - A )^2 - \mathbb{E}[ -25 ( f_{\mathrm{opt}}(X) - A )^2 |X] 
 = -25 \{ A^2 -2A f_{\mathrm{opt}}(X) + 2 f_{\mathrm{opt}}(X) -4/3  \}$ (where the expectation was evaluated with respect to the distribution of $A$). 
 This function $g$ satisfies  the conditional mean zero constraint (\ref{ortho.constr}). 
On the other hand, 
under SIMSL (\ref{the.truth}), 
the term $\mu$   is 
 $\mu(X) := 8 + 4 X_1 - 2 X_2 - 2X_3 + \mathbb{E}[- 25 ( f_{\mathrm{opt}}(X) - A )^2 |X] 
 = 
 8 + 4 X_1 - 2 X_2 - 2X_3 -25 \{f_{\mathrm{opt}}^2(X)  - 2 f_{\mathrm{opt}}(X) + 4/3  \}$, 
which corresponds to the $X$ ``main'' effect (this does not involve the variable $A$). 
Given $X$,  the term 
  $g(\beta^\top X,A)$ as a function of $A$ is maximized at $A = f_{\mathrm{opt}}(X)$, implying  that 
  $f_{\mathrm{opt}}(X)$ of Scenario 1 
corresponds to the optimal individualized dose rule specified in  (\ref{f.opt}). 
 We  can similarly formulate  Scenarios 2, 3 and 4, however,   in these cases,  
the $g$ term in  SIMSL (\ref{the.truth})   is misspecified
  (i.e.,  the term $g$ cannot be expressed in terms of a single-index $\beta^\top X$). Thus, 
  a more general function $g(X, A)$ (which is subject to a more general identifiability condition $\mathbb{E}[g(X,A)|X] =0$),  instead of  the single-index function $g(\beta^\top X,A)$, 
  should be employed  for  the heterogeneous dose effect $g$. 
 In such scenarios, 
 SIMSL (\ref{the.truth}), with optimization (\ref{sub.problem2}), provides the optimal single-index-based approximation to $g(X,A)$ (with respect to the KL divergence, see (\ref{glm.simsl.criterion}).   
  \end{Remark}

Following \cite{Chen.IDR}, we set $p=30$ in Scenario 1, and  $p=10$ for Scenarios 2, 3 and 4.  
For each simulated dataset, we apply the proposed method 
of estimating the $A$-by-$X$ interaction term in the SIMSL (\ref{the.truth}) 
 based on  (\ref{sub.problem2}),   
and  the optimal dose rule  
$f_{\mathrm{opt}}$ by 
$\hat{f}(X) 
= \underset{a \in \mathcal{A}}{\text{argmax}} \  \hat{g}^\ast(\hat{\beta}^{\ast \top} X, a)$.   
We simulated 200 data sets for each scenario. 
For comparison, we report results 
of the estimation approaches considered in \cite{Chen.IDR}, 
including their Gaussian kernel-based outcome-weighted learning (K-O-learning) and linear kernel-based  outcome-weighted learning (L-O-learning).  We also report a support vector regression 
\citep[SVR;][]{Vapnik.SVR, Smola.SVR}  
with a Gaussian kernel to estimate the 
nonlinear relationship between $Y$ and $(A, X)$ \citep{Zhao.reinforcement.learning} that was used for comparison. 
In Scenario 1, we used $(A, X)$ 
as the predictors for the outcome in the SIMSL. 
In Scenarios 2, 3 and 4, we used $(A, X, X^2)$ (i.e., including a quadratic term in $X$)  as the predictors for the  SIMSL. 

Since we are simulating data from known models in which the true relationship $\mathbb{E}[Y |X, A]$ 
is known, we can compare the estimated dose rules $\hat{f}$ derived from each method in terms of the value (\ref{value.function}). 
Specifically, an independent 
test set of size  $\tilde{n}=5000$ was generated and the value (\ref{value.function}) of $\hat{f}$ was approximated using 
$\hat{\mathcal{V}}(\hat{f}) = \tilde{n}^{-1} \sum_{i=1}^{\tilde{n}} \mathbb{E}[Y_i | X_i, A_i=\hat{f}(X_i)]$, 
for each simulation run. 
Given each scenario and a training sample size $n$, 
we replicate the simulation experiment  $200$ times, each time estimating the value. 
Again, following \cite{Chen.IDR}, 
we report the averaged estimated values (and standard deviations) 
for the cases where $\hat{f}$  
is  estimated from a training set of size $n=50, 100, 200, 400$ and $800$ for  Scenario 1 and 2, 
and the cases with 
$n=200$ and $800$ for Scenario 3 and 4.  
The simulation results are given in Table~\ref{table1} and \ref{table2}.

\begin{table}
    \begin{tabular}{llllll}
    ~          & n   & SIMSL        & K-O-learning & L-O-learning & SVR \\ \hline
    Scenario 1 & 50  &  1.04 (4.06) & 4.78 (0.48)  & \bf{4.83 (1.40)}  & -12.21 (7.53) \\
    ~          & 100 &  \bf{6.63 (0.63)} & 5.69 (0.40)  & 5.39 (0.93) & -2.57 (6.34)  \\
    ~          & 200 &  \bf{7.45 (0.20)} & 6.68 (0.26)  &6.85 (0.34) &  3.46 (1,97) \\
    ~          & 400 &  \bf{7.77 (0.08)} & 7.28 (0.15)  & 7.41 (0.14) &  6.13 (0.47) \\
    ~          & 800 & \bf{7.88 (0.04)} & 7.54 (0.08)  & 7.67 (0.08) & 7.36 (0.12) \\
    Scenario 2 & 50  & 0.90 (2.04) & \bf{2.00 (0.29)}  & 1.16 (0.71) &  -1.96 (1.70) \\
    ~          & 100 &  \bf{3.65 (0.76)} & 2.19 (0.43)  & 1.57 (0.52) & 0.24 (1.42) \\
    ~          & 200 &  \bf{4.71 (0.41)} & 2.84 (0.37)  & 2.02 (0.30) & 2.01 (0.84) \\
    ~          & 400 &  \bf{5.25 (0.20)} & 3.69 (0.27)  & 2.30 (0.18) & 3.47 (0.37) \\
    ~          & 800 &  \bf{5.59 (0.12)} & 4.41 (0.19)  & 2.49 (0.10)& 4.35 (0.19) \\ \hline
    \end{tabular}
        \caption {Average (and sd) value $\hat{\mathcal{V}}(f)$ from 200 replicates from the randomized trial scenarios. In both settings, the oracle $f_{\mathrm{opt}}$ 
        attains a value function $\mathcal{V}(f_{\mathrm{opt}} ) = 8$  
        (boldface denotes the largest in each row).
}
\label{table1}
\end{table}


\begin{table}
    \begin{tabular}{llllll}
    ~          & n   & SIMSL        & K-O-learning & K-O-learning(Prp) & SVR          \\ \hline
    Scenario 3 & 200 & \bf{4.03 (0.97)}  & 2.68 (0.30)  & 2.74 (0.29)     & 1.99 (0.83)  \\
    ~          & 800 &  \bf{5.46 (0.20)} & 4.06 (0.30)  & 4.19 (0.20)     & 4.09 (0.28)  \\ \hline
    Scenario 4 & 200 &  \bf{4.07 (0.72)} & 3.29 (0.28)  & 3.23 (0.28)     & -0.95 (1.57) \\
    ~          & 800 &  \bf{5.51 (0.19)} &  4.91 (0.14)  & 4.73 (0.17)     & 3.04 (0.52)  \\
    \end{tabular}
            \caption {Average (and sd) value $\hat{\mathcal{V}}(f)$ from 200 replicates from observational studies. In both settings, the oracle $f_{\mathrm{opt}}$ 
        attains a value function $\mathcal{V}(f_{\mathrm{opt}} ) = 8$ (boldface denotes the largest in each row). 
} \label{table2}
\end{table}

The results in Table~\ref{table1} and \ref{table2} 
indicate that the proposed regression method for optimizing individualized dose rules outperforms the alternative approaches presented in \cite{Chen.IDR} in all cases except when the training sample size is very small $(n=50)$. In Table~\ref{table2}, K-O-learning(Prp) refers to the propensity score-adjusted K-O-learning of \cite{Chen.IDR}. 
When the sample size is very small, the outcome-weighted learning approaches outperform 
the regression-based approaches (i.e., SIMSL and SVR), especially for Scenario 1 (with $p=30$) where the regression approaches exhibit large variances. 
However, when $n = 100$, 
the performance of the SIMSL approach improves dramatically in terms of both value and small variance. 
We also note that using $(A, X, X^2)$  instead of $(A, X)$ as predictors of the SIMSL 
 in Scenario 2 lead to a substantial improvement in performance. If  $(X, A)$ is used for the SIMSL in Scenario 2,   the estimated values (and sd) are: 
  $-0.98 (2.15),  0.56 (1.54), 1.91 (0.89),  2.70 (0.64)$ and  $3.23 (0.41)$, 
for sample sizes $n=50, 100, 200, 400$ and  $800$, respectively.

\section{Application to optimization of the warfarin dose with clinical and pharmacogenetic data}

In this section, the utility of the SIMSL approach to personalized dose finding is illustrated from an anticoagulant study.  
Warfarin is a widely used anticoagulant to treat and prevent blood clots. The therapeutic dosage of warfarin varies widely across patients and its administration must be closely monitored to prevent adverse side effects.  
Our analysis of the data will broadly follow that of \cite{Chen.IDR}.  
After removing patients with missing data, the dataset provided by \cite{Warfarin.dataset} (publicly available to download from https://www.pharmgkb.org/downloads/) 
consists of 1780 subjects, 
 including information on patient covariates $(X)$, final therapeutic dosages $(A)$, and patient outcomes (INR, International Normalized Ratio). 
 INR is a measure of how rapidly the blood can clot. 
For patients prescribed warfarin, the target INR is around 2.5. In order to convert the INR to a measurement responding to the warfarin dose level, we construct an outcome $Y= - |2.5 -  \mbox{INR} |$, and a larger value of $Y$ is considered desirable.

There were 13 covariates $X = (X_1,\ldots, X_{13})^\top$ in the dataset (both clinical and
pharmacogenetic variables):
height ($X_1$), weight  ($X_2$), age  ($X_3$), use of the cytochrome P450  enzyme inducers ($X_4$; the enzyme inducers considered in this analysis includes phenytoin, carbamazepine, and rifampin), use of amiodarone ($X_5$), gender ($X_6$; 1 for male, 0 for female), African or black race ($X_7$), Asian race ($X_8$), the VKORC1  A/G genotype ($X_9$), the VKORC1 A/A genotype ($X_{10}$), the CYP2C9  *1/*2 genotype ($X_{11}$), the CYP2C9 *1/*3 genotype ($X_{12}$), and the other CYP2C9 genotypes (excluding the CYP2C9 *1/*1 genotype  which is taken as the baseline genotype) ($X_{13}$). Further details on these covariates are given in \cite{Warfarin.dataset}.   The first 3 covariates (height, weight, age) were treated as continuous variables,  and we standardized them to have mean zero and unit variance; the other 10 covariates are indicator variables.

In estimating the optimal individualized dose rule $f_{\mathrm{opt}}$, 
modeling the drug (dose level $A$) interactions with the patient covariates $X$ in their effects on  $Y$ is essential. 
Under the proposed SIMSL approach (\ref{the.truth}), 
$f_{\mathrm{opt}}(X) =  \  \underset{a \in \cal{A}}{\text{argmax}} \ g(\beta^\top X, a)$ 
and thus 
 the $A$-by-$X$ interaction effect term $g(\beta^\top X, A)$ 
is the target component of interest. 
In SIMSL, due to orthogonality (\ref{ortho}), modeling the potentially complicated function $\mu$ in (\ref{the.truth}) can be avoided in the estimation of $(g, \beta)$. However, modeling the $X$ main effects (i.e., $\mu(X)$) generally improves  the estimation efficiency (i.e., giving smaller variances for estimators; see \cite{CSIM, MC} for the theoretical justification for the case where treatment $A$ is a discrete/binary variable) for $(g, \beta)$ and thus that for $f_{\mathrm{opt}}$ (see Supplementary Materials Section C.1 for a simulation illustration where the $X$ main effect is incorporated to the estimation of $f_{\mathrm{opt}}$).

\begin{figure}  
\includegraphics[width=6.6in, height = 2.4in]{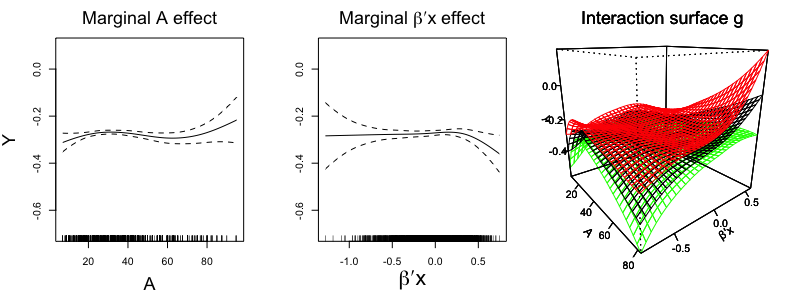} 
\caption[]{ 
The first two panels: 
the marginal effect of dose $A$ (left panel) and  that of the estimated single-index  (middle panel) with 95\% confidence bands (dashed curves) given the estimated  $\beta^\top X$.   
The third panel: the estimated link surface $(g)$  for the dose ($A$) and index ($\beta^\top X$)  interaction; 
the red and green surfaces are at $\pm 2 $  standard error from the estimated  surface (the black) in the middle, conditioning on the estimated single-index.  
 } \label{simsl_warfarin}
\end{figure}

 In this  application, 
we model the unspecified component $\mu(X)$ of (\ref{the.truth})  with an additive working model, which  
consists of a set of linear terms for the indicators $X_{4}, \ldots, X_{13}$ 
and a set of cubic $P$-spline smooth terms for the continuous covariates $X_1$, $X_2$ and $X_3$. 
These terms are estimated alongside the heterogeneous does effect term  $g(\beta^\top X, A)$  by the the procedure described in  Supplementary Materials  Section C.1, which is a slight modification of that in Section~\ref{sec.estimation.algo}.  
 We focus on the component $g(\beta^\top X, A)$ in (\ref{the.truth}).  
The estimated $\beta$  is  $(0.18, 0.02, -0.02, 0.52, -0.41, 0.10, -0.23,  0.32,-0.11, 0.05, 0.02,-0.34,-0.48)^\top$. 
The fitted $\beta$ (and its bootstrap confidence interval) obtained without incorporating the $X$ main effects is provided in Supplementary Materials Section D. 
The third panel  in Figure~\ref{simsl_warfarin} displays 	
the estimated interaction surface plot of the  $2$-dimensional  surface-link function $g(\beta^\top X, A)$, 
showing an interactive relationship on the index-treatment domain.   
    The first two panels in 
Figure~\ref{simsl_warfarin} display the plots for the estimated  marginal  effect function for the dose $A$ and that for the estimated single-index $\beta^\top X$.

We construct a 95\% normal approximation bootstrap confidence interval  for $\beta$, based on 500 bootstrap replications (see Supplementary Materials Section  for the constructed confidence intervals and for a coverage probability simulation).  The confidence intervals for the $\beta_j$'s associated with the covariates height ($X_1$), the use of the cytochrome P450  enzyme inducers ($X_4$), the use of amiodarone ($X_5$), the CYP2C9 *1/*3 genotype ($X_{12}$), and the other CYP2C9 genotypes ($X_{13}$) do not include $0$. We infer that these covariates are potentially clinically important drug effect modifiers, interacting with warfarin in their effects on INR.

\cite{Chen.IDR} noted that the analysis results from \cite{Warfarin.dataset}, as well as their linear kernel-based outcome-weighted learning results, suggest  increasing the dose if patients are taking Cytochrome P450 enzyme ($X_4$). Roughly speaking, the interaction surface $g$ (the right-most panel) in Figure~\ref{simsl_warfarin} indicates that for a larger value of  $\beta^\top X$  (e.g., $\beta^\top X > 0$), a relatively high dose $A$ (e.g., $A > 50$) may be preferred,  whereas for a smaller value of $\beta^\top X$ (e.g., $\beta^\top X < -0.5$), a moderate or a relatively low  dose $A$  (e.g., $A < 50)$ may be preferred.  Considering the sign of the estimated coefficient  ($\hat{\beta}_4 = -0.58$) associated with $X_4$,  this is roughly consistent with  \cite{Warfarin.dataset} and \cite{Chen.IDR}. 

To evaluate the performance of the individualized dose rules estimated from the $6$ methods, including the propensity score-adjusted 
outcome-weighted learning with a linear/Gaussian kernel, denoted as L-O-learning(Prp) and K-O-learning(Prp), respectively) considered in Section~\ref{sec.simulation}, 
we randomly split the dataset at a ratio of 1-to-1 into a training set and a testing set, replicated 100 times, each time estimating $f_{\mathrm{opt}}$  
using the $6$ methods based on the training set, and estimating the value (\ref{value.function}) of each estimated $f_{\mathrm{opt}}$ based on the testing set.  
Unlike the simulated data  in Section~\ref{sec.simulation}, 
the true relationship between the covariate-specific dose and the response is unknown.  
Therefore, for each  dose rule $f$, we need to estimate the value  (\ref{value.function}) from the testing data.  Given a dose rule $f$, 
 only a very small proportion (or none) of the observations will satisfy $A_i = f(X_i)$, and thus 
only a very small proportion (or none) of the observations in the testing data will contribute information to estimate the value (\ref{value.function}). 
However, 
\cite{commentCai2016} noted that the value  (\ref{value.function}) for each $f$ can also be written as 
$\mathcal{V}(f) = \mathbb{E}[ \mathbb{E}[Y | A=f(X), f(X) ] ]$. 
Therefore, 
using a $2$-dimensional smoother  of $A$ and $f(X)$ for $Y$, 
one may first obtain a nonparametric estimate of 
$\mathbb{E}[Y | A, f(X) ]$,  
 denoted as $\hat{m}({A,f(X)})$, 
and then $\mathcal{V}(f)$
 may be estimated as $\hat{\mathcal{V}}(f)  = n^{-1} \sum_{i=1}^n \hat{m}({f(X_i),f(X_i)})$. 
Specifically, 
given a dose rule $f$ estimated from a training set, 
we can estimate 
$\mathbb{E}[Y | A, f(X) ] $ based on $(Y_i, A_i, f(X_i))$ from a test set, 
using a set of thin plate regression spline bases obtained from a rank-100 eigen-approximation to a thin plate spline, with the smoothness parameter  selected by REML, implemented via the R \citep{R} function \texttt{mgcv::gam} \citep{mgcv}. 
A thin plate spline is an isotropic smooth; isotropy is often appropriate for  two variables observed on the same scale, which is the case here.

\begin{figure}   
\includegraphics[width=5.4in, height = 2.7in]{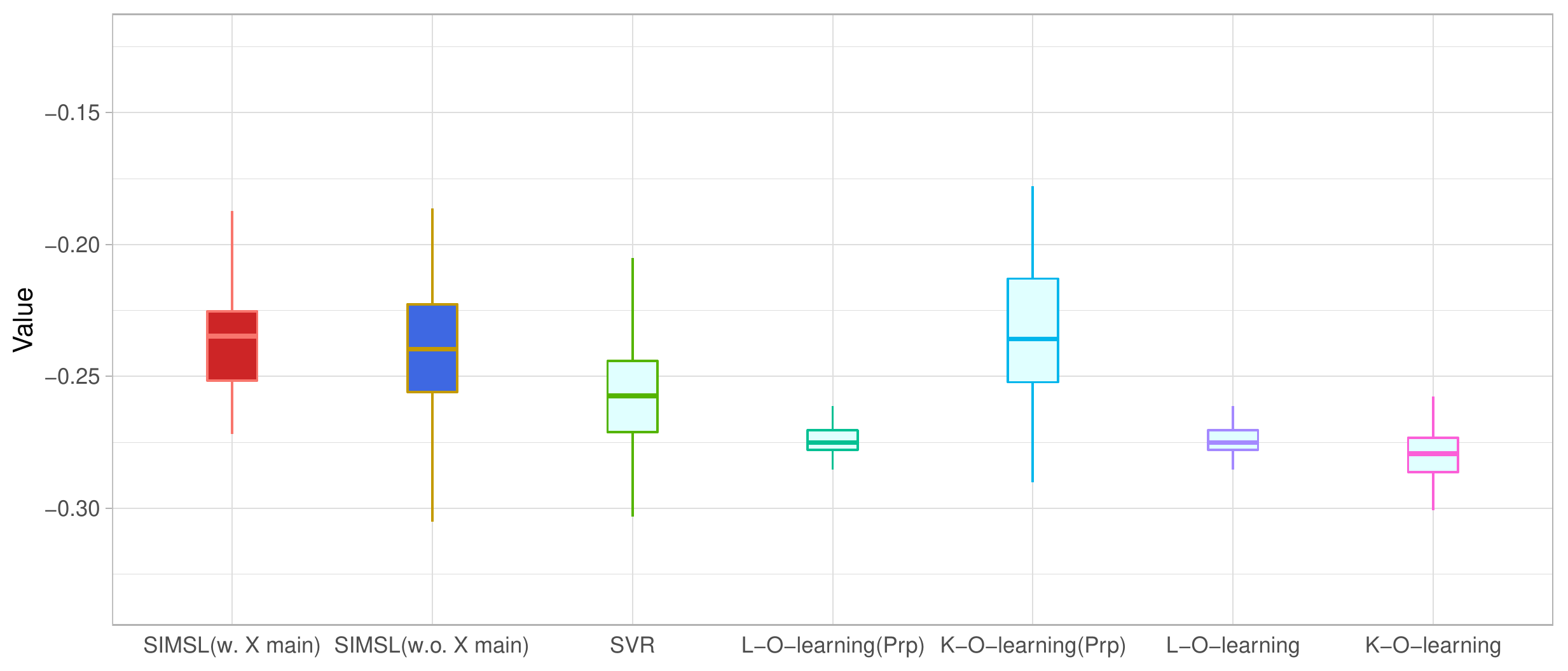} 
\vspace*{-5pt}
\caption[]{ Boxplots of the estimated values  of the individualized dose rules using $7$ approaches, obtained from $100$ randomly split testing sets. 
Mean (and standard deviation) of the value estimates: 
SIMSL(w. $X$ main): -0.233 (0.03); SIMSL(w.o. $X$ main): -0.237 (0.03); SVR: -0.256 (0.02);  L-O-learning(Prp): -0.274 (0.01); 
K-O-learning(Prp): -0.234 (0.03); L-O-learning: -0.274 (0.01); K-O-learning: -0.279 (0.01).} \label{simsl_warfarin_value}
\end{figure}

Figure~\ref{simsl_warfarin_value} displays a boxplot describing the distributions for the estimated values (\ref{value.function}) 
of  ``SIMSL(w. $X$ main)''  (SIMSL  incorporating $\mu$ in the estimation) and ``SIMSL(w.o. $X$ main)'' (SIMSL without  incorporating $\mu$ in the estimation), 
and the other five estimation methods described in Section \ref{sec.simulation}, obtained from the aforementioned $100$ random training/testing splits. 
The boxplots indicate that the proposed SIMSL methods  
(note that SIMSL(w. $X$ main) slightly outperforms SIMSL(w.o. $X$ main)) 
and the propensity-score adjusted K-O-learning of \cite{Chen.IDR} perform at a similar level, while outperforming all other approaches. 
The results illustrate  the potential utility of the proposed regression approach to optimizing individualized dose rules. 
In comparison to the outcome-weighted learning  approach of  \cite{Chen.IDR},  
one advantage of the proposed approach is that it allows visualization of the estimated 
interactive structure on the dose-index domain as illustrated in the right panel of Figure 3. Additionally, 
if each of the covariates is standardized to have, say, unit variance, then 
the relative importance of each covariate  in characterizing the heterogeneous dose response can be determined by the magnitude of the estimated coefficients in $\beta$, 
rendering a potentially useful interpretation  when examining the drug-covariates interactions.

\section{Discussion}\label{sec.discussion} 

In this paper, we proposed a variant of a single-index model that utilizes a surface link-function as a function of a linear projection of covariates and a continuous ``treatment''  variable.  This single index model with a surface link can effectively estimate the effect on a response of possibly nonlinear interactions between a set of covariates and the treatment variable defined on a continuum. The proposed regression model  is useful for developing personalized dose rules in precision medicine, and more generally, in a situation where we are particularly interested in modeling interactions between a vector of covariates and a real-valued predictor of interest. The model gives an intuitive method for modeling such interactions, without the need for a significant change in the established generalized additive regression modeling framework. 

One important limitation is that the confidence band associated with the estimated surface $g$ is computed conditional on the estimated $\beta^\top X$, and the uncertainty in $\beta$ is not accounted for. The fact that  the domain of $g$ varies depending on the estimate of $\beta$ complicates the confidence band construction for $g$. One potential 
solution would be to consider a Bayesian approach and use a posterior distribution of $g(\beta^\top X, A)$, and    
 make probabilistic statements about  the prediction given $(X, A)$.  

Depending on context, for scientific interpretability of the model, it may be sometimes desirable to consider shape constraints such as monotonicity or convexity/concavity (see Supplementary Materials Section C.4 for some discussion on this topic). The development of a Bayesian framework for this regression model, with potential monotonicity or convexity/concavity constraints on the link surface is currently under investigation. 
In many applications, only a subset of variables may be useful in determining an optimal individualized dose rule.  Also, high-dimensional settings can lead to instabilities and issues of overfitting. Forthcoming work will introduce a regularization method that can both avoid overfitting 
and choose among multiple potential covariates by obtaining a sparse estimate of the single-index coefficient $\beta$.  
Future extensions of this work could also include an extension 
to incorporate a functional covariate.

\bigskip
\begin{center}
{\large\bf SUPPLEMENTAL MATERIALS}
\end{center}

\begin{description}

\item[Supplementary Materials:] a pdf file containing the proof of Proposition~1, a data analysis example and additional simulations illustrating an application of the generalized single-index regression approach described in Section~\ref{sec.generalized}.

\item[R-package for SIMSL routine:]  R-package \texttt{simsl} \citep{simsl.package} available on CRAN containing code to perform the proposed single-index  regression method, and the datasets and the simulation examples illustrated in this article. For the case of discrete-valued treatments,  we refer to R-package \texttt{simml} \citep{simml.package}. 

\end{description}

{\small 
\bibliography{refs}
}

\end{document}